\documentclass{article}
\usepackage{arxiv}

\usepackage{url}
\usepackage{cite}
\usepackage{amsmath,amssymb,amsfonts}
\usepackage{algorithmic}
\usepackage{graphicx}
\usepackage{textcomp}
\usepackage{subfigure}
\usepackage{multirow,array}
\usepackage{tikz}
\usetikzlibrary{shapes,arrows}
\usepackage{pgfplots}
\usepackage{threeparttable}
\usepackage{cite}
\pgfplotsset{compat=1.14}
\PassOptionsToPackage{hyphens}{url}\usepackage{hyperref}

\usepackage[lined,linesnumbered,ruled]{algorithm2e}
\usepackage{lscape}
\usepackage{rotating}
\usepackage{wrapfig,lipsum,booktabs}
\usepackage{color}
\usepackage[Symbol]{upgreek}
\usepackage{bm}
\usepackage{mathrsfs}
\usepackage{diagbox}

\usepackage{url,times, tabularx}
\usepackage{units}
\usepackage{balance}
\usepackage{soul}
\usepackage{cleveref}

\usepackage{xcolor}

\title{Sensing of inspiration events from speech: comparison of deep learning and linguistic methods}

\author{Aki H\"arm\"a$^1$\thanks{Currently with DACS, Maastricht University, The Netherlands. $^1$Philips Research, Eindhoven, The Netherlands. $^2$Philips Research, Bangalore, India} ,  Ulf Grossekath\"ofer$^1$, Okke Ouweltjes$^1$ and Venkata Srikanth Nallanthighal$^2$
}

\begin{document}

\maketitle

\begin{abstract}
Respiratory chest belt sensor can be used to measure the respiratory rate and other respiratory health parameters. Virtual Respiratory Belt, VRB, algorithms estimate the belt sensor waveform from speech audio. In this paper we compare the detection of inspiration events (IE) from respiratory belt sensor data using a novel neural VRB algorithm and the detections based on time-aligned linguistic content. The results show the superiority of the VRB method over word pause detection or grammatical content segmentation. The comparison of the methods show that both read and spontaneous speech content has a significant amount of ungrammatical breathing, that is, breathing events that are not aligned with grammatically appropriate places in language. This study gives new insights into the development of VRB methods and adds to the general understanding of speech breathing behavior. Moreover, a new VRB method, VRBOLA, for the reconstruction of the continuous breathing waveform is demonstrated.
\end{abstract}
\noindent\textbf{Index Terms}: 
Pathological speech sensing, speech breathing, respiratory health and fitness, automatic speech recognition

\section{Introduction}
In recent year we have seen increase in speech-based health sensing methods \cite{cummins_speech_2018}. Often, the goal is to detect a condition, such as a respiratory infection \cite{albes_squeeze_2020}, or a specific diagnostic condition such as OSA \cite{goldshtein_automatic_2011}, Alzheimer's \cite{haider2019assessment,LuzHaiderEtAl20ADReSS} or Parkinson's disease \cite{moro2020using}. One may also aim at estimating continuous values such as the age of a talker \cite{ghahremani2018end}, or simultaneous physiological sensor signal from speech. For example, it would be very useful to be able to estimate the blood sugar level of a diabetic patient \cite{sidorova_blood_2022} or respiratory parameters of a talker \cite{nallanthighal_deep_2019} directly from the speech of a caller, for example, in a tele-health application. The current paper studies the problem of respiratory health sensing from speech using the VRB method. The Virtual Respiratory Belt, VRB, processing uses speech audio to model a signal captured using a chest-worn Respiratory Inductance Plethysmography, RIP, belt sensor. RIP measurements can be used to get an estimate of the respiratory parameters such as respiratory rate and tidal volume \cite{heyde_respiratory_2014,laufer_tidal_2019}. McKenna {\it et al} demonstrated that it is also possible to predict, from RIP sensor data, the values of spirometry measurements during {\it speechlike} breathing \cite{mckenna_accuracy_2019}. 

Some of the early VRB work \cite{nallanthighal_deep_2019} was done using log-Mel spectrum data and RNN neural networks. A VRB task was also included in the 2020 Paralinguistic challenge \cite{schuller_interspeech_2020}. Various CNN network architectures have been recently found useful in the parallel problem of covid-19 detection \cite{effati_performance_2023} and it was shown in a recent paper [anon] that the performance of the VRB modeling can be further improved by using pre-trained transformer networks such as Hubert \cite{hsu_hubert_2021} or Whisper \cite{radford_robust_2022}. In the current paper one such design is introduced, and we demonstrate its performance in the task of the detection of inhaling, or Inspiration Events, IEs, from read and spontaneous speech content. In addition, we introduce and demonstrate the performance of a novel modification, VRBOLA, of the conventional VRB method, which is using overlap-add processing in the waveform reconstruction. 

{\it Ingressive phonation}, that is, speech vocalization during inspiration, is relatively rare \cite{eklund_pulmonic_2008}, and the inspiration events in typical speech occur in pauses between words. Speech pauses can be detected using speech activity detection, SAD, algorithms. It is shown in this paper that by a selection of long pauses in speech using a good SAD can give a rough estimate of the occurrence of breathing events in speech. 

There is a complex relationship between spoken language and our respiratory behavior influenced by breathing planning \cite{winkworth1994variability,wlodarczak_respiratory_2017,egorow_employing_2019,fuchs2021respiratory} which makes it challenging to estimate the VRB signal directly from speech, see discussion in [anon]. The breathing planning is expected to be guided by the rules of language so that IEs are placed where there is a natural pause in speech, for example, at end of a phrase or a sentence. In fact, some studies on read speech respiration have found that IEs in speech occur {\sl almost exclusively at grammatically appropriate places}\cite{winkworth1994variability}. In conversational speech IEs occur also elsewhere, e.g., upto 13 \% in \cite{wang_breath_2010}. One could propose to use the detection of grammatically appropriate places in speech as another method to detect IEs. In this paper, the idea is tested and compared against the VRB method. 

 This paper compares the performance of the proposed VRB algorithm to a method based on detection of speech pauses, and a method based on detection of {\it grammatically appropriate} places in speech, respectively. As expected, the VRB has the best performance in the detection of IEs in both read and spontaneous speech. The analysis of the results gives very interesting insights about speech breathing and design of algorithms for speech-based respiratory sensing. It also seems to show that, in the data used in this study, the assumption of {\it grammatical breathing} does not appear very strong but many IEs in both read and spontaneous speech are examples of {\it ungrammatical} breathing, i.e., inspiration events occurring in other parts of speech. Ungrammatical breathing occurs in both read speech and spontaneous speech but is more prevalent in spontaneous speech. 

\section{Breathing waveform estimation}
\begin{figure}
    \centering
    \includegraphics[width=12cm]{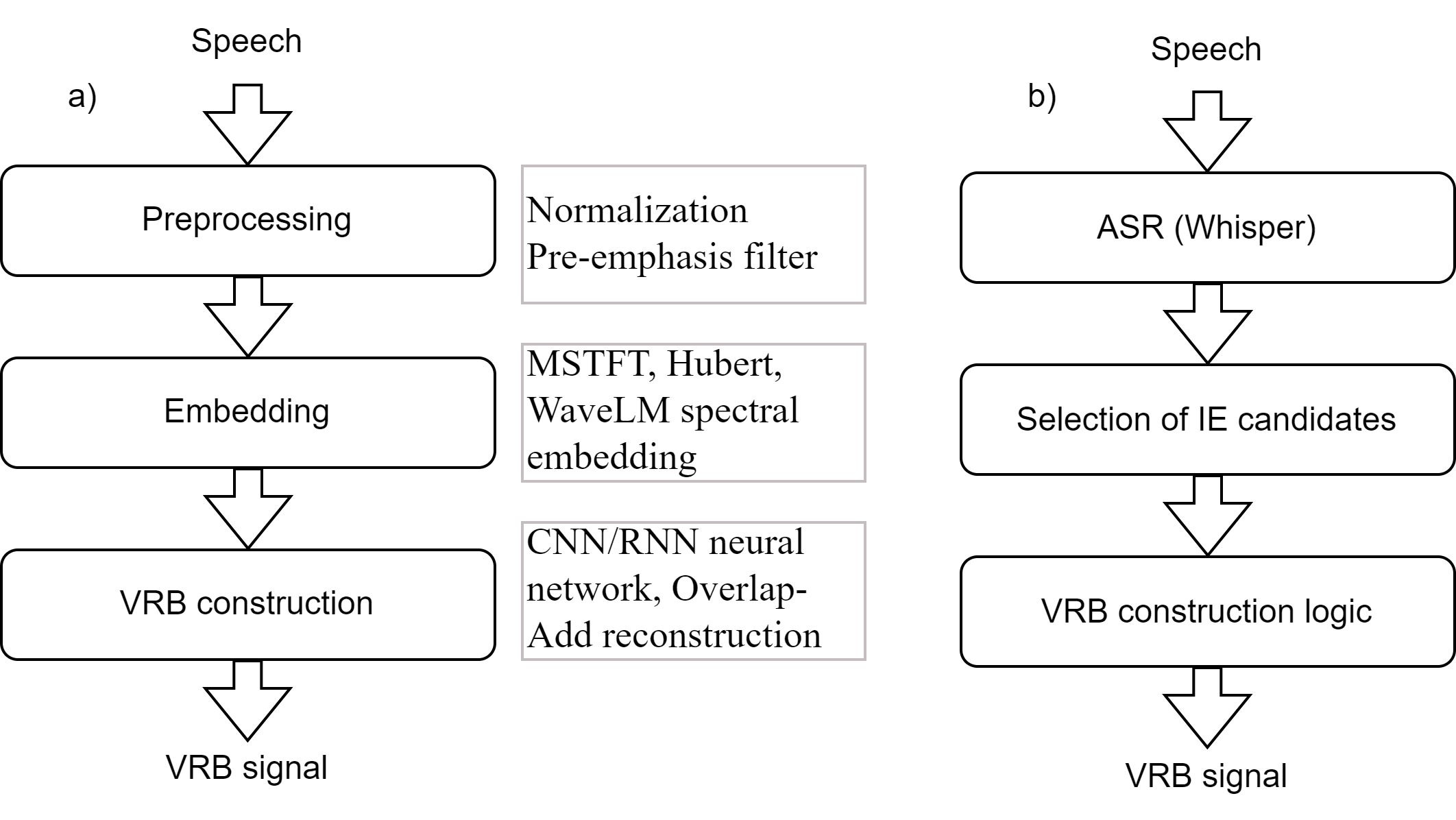}
    \caption{Block diagram for breathing waveform estimation from speech using a) VRB b) ASR and logic}
    \label{fig:block}
\end{figure}

The basic block diagram of a basic VRB algorithm is shown in Fig.\,\ref{fig:block}a. The speech preprocessing block contains typical steps of signal level normalization by automatic gain control, pre-emphasis filtering to remove low frequency noises and to flatten the spectrum, and possibly an insertion of a background noise floor to make the models more robust for background noise conditions. In the experiments reported in \cite{nallanthighal_deep_2021} 
the signal was transformed to a Log-Mel-spectrogram representation. In [anon], it was demonstrated that the performance of both instantaneous breathing waveform modeling and forecasting breathing of events is significantly improved by using large pre-trained wave2vec transformer models, such as Hubert \cite{hsu_hubert_2021}, in the preprocessing of the data before the neural network. 

\subsection{VRB wave modeling}
The experiments of the current paper use a VRB model architecture that gave the highest Pearson correlation in another study \cite{nallanthighal_deep_2021} between the obtained VRB signal and the real belt signal. This model uses Hubert as the embedding model to generate vectors corresponding to 20ms processing windows at the sample rate of 16kHz. The vectors are stacked to a a matrix of 256 columns which are further processed using a 3-layer Gated Recurrent Unit, GRU, network with 64 hidden units, and finally one dense linear layer to produce $K$ samples of the VRB signal corresponding to the speech signal. 

\begin{algorithm}
\caption{Frame-based VRB modeling}\label{alg:vrb}
\begin{algorithmic}[1]
\STATE Use pretrained Hubert model to embed the content in 30ms windows to $D=1024$-tap vectors
\STATE Stack $K=256$ vectors into a matrix ${\bf M}_{K\times D}$.
\STATE Train a cascade of a LSTM network and 1D CNN network to produce a $p$th frame of a VRB signal $b_p(k), k=0,...K-1$.
\end{algorithmic}
\end{algorithm}

The neural network was trained using the data from the dataset A introduced in Section \ref{sect:data}. The model was optimized using the smooth L1 loss function, and the model optimization was performed by the Stochastic Gradient Descent, SGD, method, both implemented in the pytorch library \cite{paszke_pytorch_2019}, respectively.

\subsection{Continuous wave reconstruction}
In previous work the VRB method reconstructed the signal sample-by-sample, or in concatenated frames of $N$ samples. It turns out that it is very useful to reconstruct the VRB signal using an overlap-add, OLA, method. In this paper, this method is called VRBOLA. The last step of Algorithm \ref{alg:vrb} produces a VRB signal fragment $b_p(k), k=0,...K-1$ for a $p^{th}$ block. In the VRBOLA method reconstruction is carried out by applying the overlap-add method using a sequence of window functions $w_p(t)$ so that the reconstruction of the final VRB signal is given by
\begin{equation}
    b(t) = \sum_{p=-\infty}^{\infty} w_p(t)b_p(t-pS).
\end{equation}
The window sequence is defined so that $w_p(t)$ is, e.g., a squared sine, in $t\in [pS,\, (p+1)S-1]$ and zero elsewhere. Overlap-add reconstruction reduces the effects of the errors that are usually larger close to the frame boundaries than in the center of the window. 

In this paper, the focus is on the detection and characterization of inspiration events, IEs. In inspiration phase the belt sensor signal typically goes up, while during speech vocalization it goes down as lungs are depleted from air. The detection of one IE in the respiration signal is defined by a local maximum followed by local minimum in the waveform. In the following experiments, the same algorithm was used for the detection of IEs from the real belt waveform and the VRB signal estimated from speech. 
\begin{algorithm}
\caption{Detection of IEs}\label{alg:cap}
\begin{algorithmic}[1]
\STATE Remove bias, sensor drift and high frequency noise by applying a third order Butterworth band-pass filter \linebreak from 0.08 Hz to 1.0 Hz using forward-backward filtering.  
\STATE Select local minima and maxima that have a minimal separation of 1s and that have normalized peak prominence that exceeds a threshold of 0.8.  
\STATE The average breathing rate is taken from the mean of the distance between the maxima.
\end{algorithmic}
\end{algorithm}

 The average non-speech inspiration and expiration durations, for example, in Chronic Obstructive Pulmonary Disease, COPD, patients are 1s and 1.7s, respectively~\cite{chatila_effects_2004}, which give a respiration rate of 0.2Hz. During speech, the average respiratory rate is approximately half of that ~\cite{wlodarczak2015breathing} and the duration of an IE is may be less than 100ms. In this paper, the sample rate of the embedding data matrix ${\bf M}$ and the modeled belt signal $b(t)$ is 50Hz, corresponding to the 20ms input size of the Hubert model.  

\section{Respiratory sensing using ASR}
In this paper, we use two databases of speech recordings with respiratory belt measurements. The dataset A has 500 subjects and it has been collected at a large medical center MAHE in India. The dataset B contains recordings of 40 talkers \cite{nallanthighal_deep_2019-1 }collected in a research lab at Philips Research in Eindhoven, the Netherlands. The VRB model used in this paper has been trained using the dataset A, and all experiments reported in the paper have been performed on dataset B. 

\begin{figure}[t]
     \subfigure(a){\includegraphics[width=7.5cm]{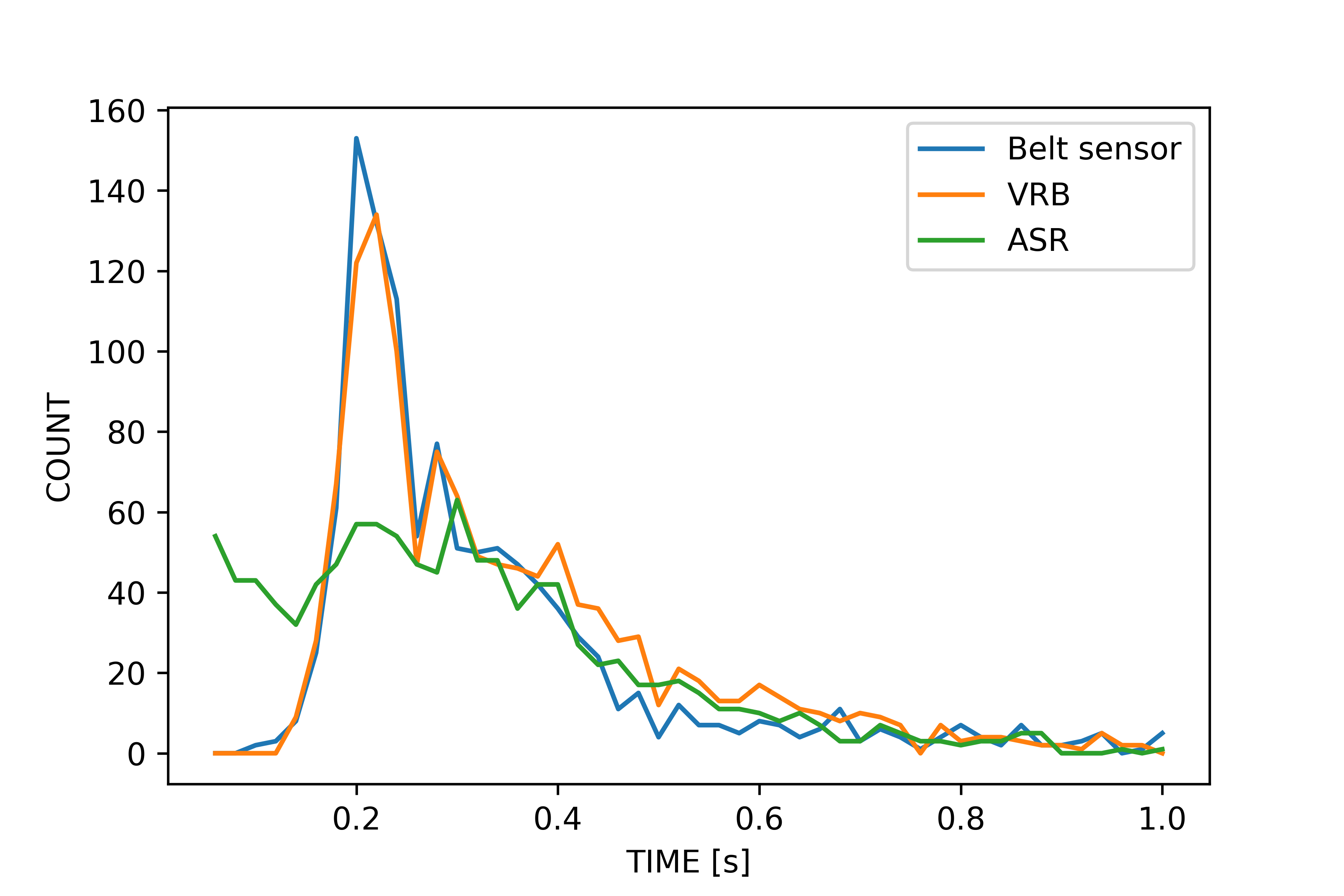}}
      \subfigure(b){\includegraphics[width=7.5cm]{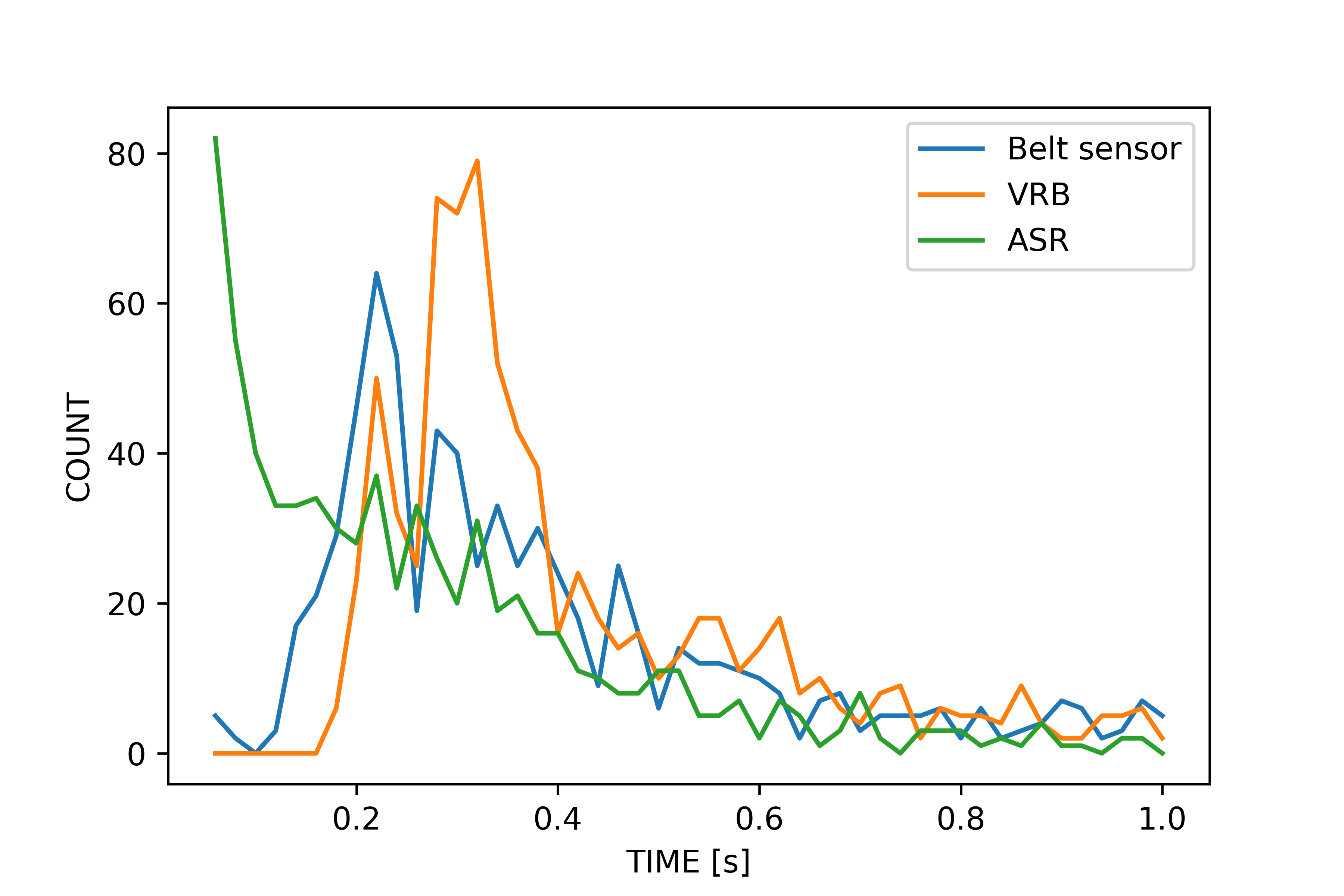}}
    \caption{Histogram of inspiration event durations based on the belt sensor data, ASR-punct, and VRB, respectively in a) read b) spontaneous speech.}
    \label{fig:hist}
\end{figure}

The database B contains forty recordings of read speech and spontaneous conversational speech. In early experiments we found that the WhisperX algorithm \cite{bain_whisperx_2022} based on the Whisper ASR \cite{radford_robust_2022} and word-level alignment works well for the data. The processing model for VRB estimation using ASR is shown in Fig.\,\ref{fig:block}b. After the ASR and word-level time-alignment the IE candidates are selected by the punctuation points in the output of WhisperX. The durations of IEs computed from the belt sensor data and the timestamps between punctuation points and the start of the next word in ASR data, and pause durations from the VRB signal, respectively, are shown in Fig. \ref{fig:hist}. The pauses in VRB follow similar statistics to the real belt measurements, which can be expected because the VRB model has been trained using similar belt data. The most common duration of an IE is 200-250ms.  

Let us define two ASR-based methods for the estimation of IEs from speech data. 
\subsection{ASR-word method} 
The underlying idea is to assume that IEs occur in long word breaks in speech. The method uses the whisperX to detect word boundaries and count all word-to-word pauses that are longer than 150ms as IE candidates. This method is based on assumption that long pauses in speech are IEs. The limit of 150ms was found optimal for the best performance in IE detection accuracy in the dataset. 

\subsection{ASR-punct method}
This method follows the observation in \cite{winkworth1994variability} that IEs usually occur at grammatically appropriate places, {\it grammatical stops}. First all ends of clauses and sentences are detected from the output of the whisperX algorithm. In most cases, the end points are indicated by punctuation. Next, the end point of each IE is selected to be the start point of the next word in the sequence after the grammatical stop. 

\section{Experiments}\label{sect:data}
 The dataset B has speech from 40 healthy volunteers in a research institute in the central Europe either reading "The Rainbow" passage \cite{fairbanks1960rainbow} or 1-2 minute free speech in English about their current project or the last holiday trip. The subjects were wearing a Respiratory Inductance Plethysmograph, RIP, belt while speaking, which measures a continuous signal corresponding to the circumference of the chest and abdomen of the subject. The VRB and VRBOLA methods were trained using the dataset A with 200 recordings collected with a medical center in India telling about a traditional fable in their local language. 

\begin{figure}[ht!]
    \centering
    \subfigure(a){\includegraphics[width=12cm]{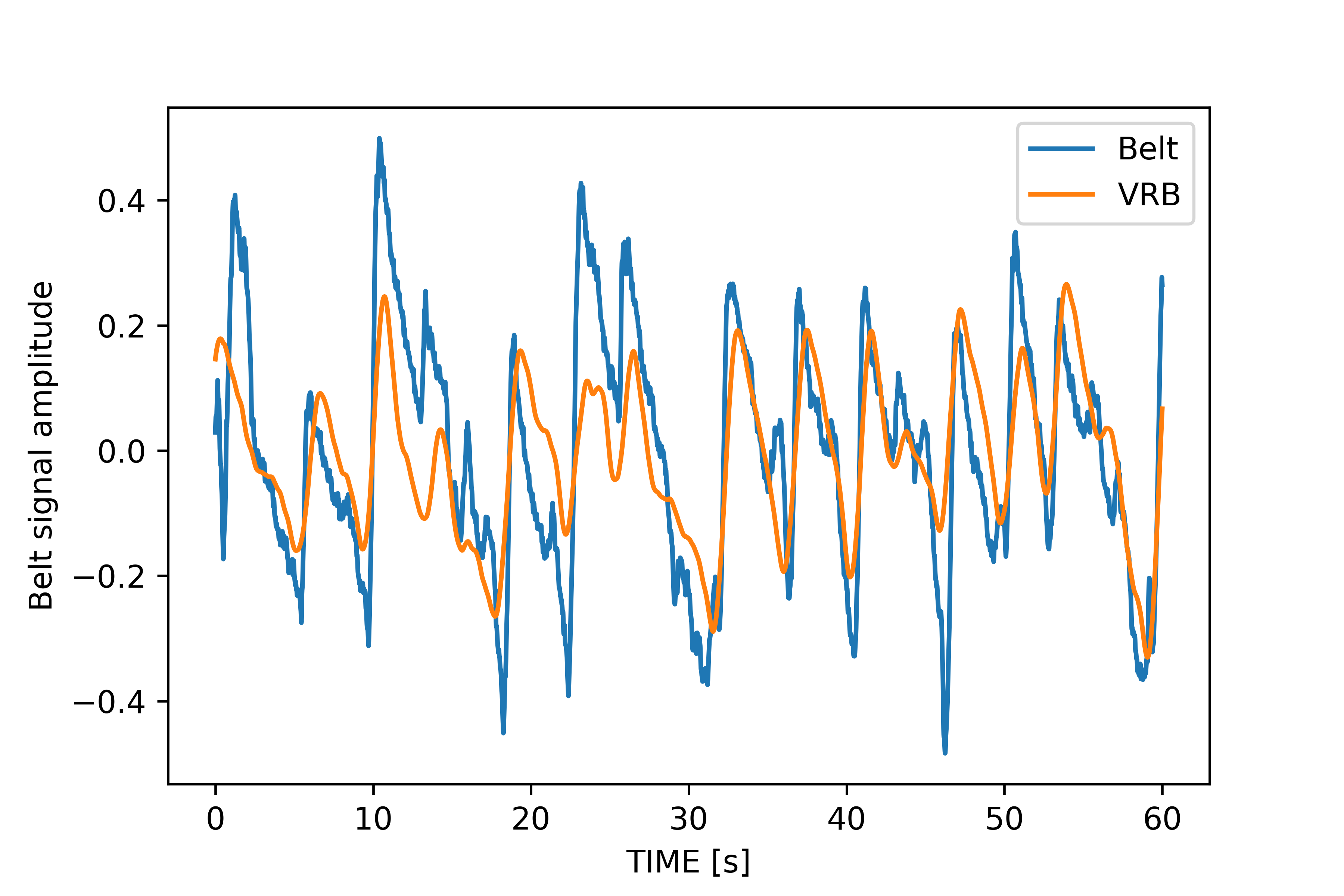}}\vspace{-8mm}
    \subfigure(b){\includegraphics[width=12cm]{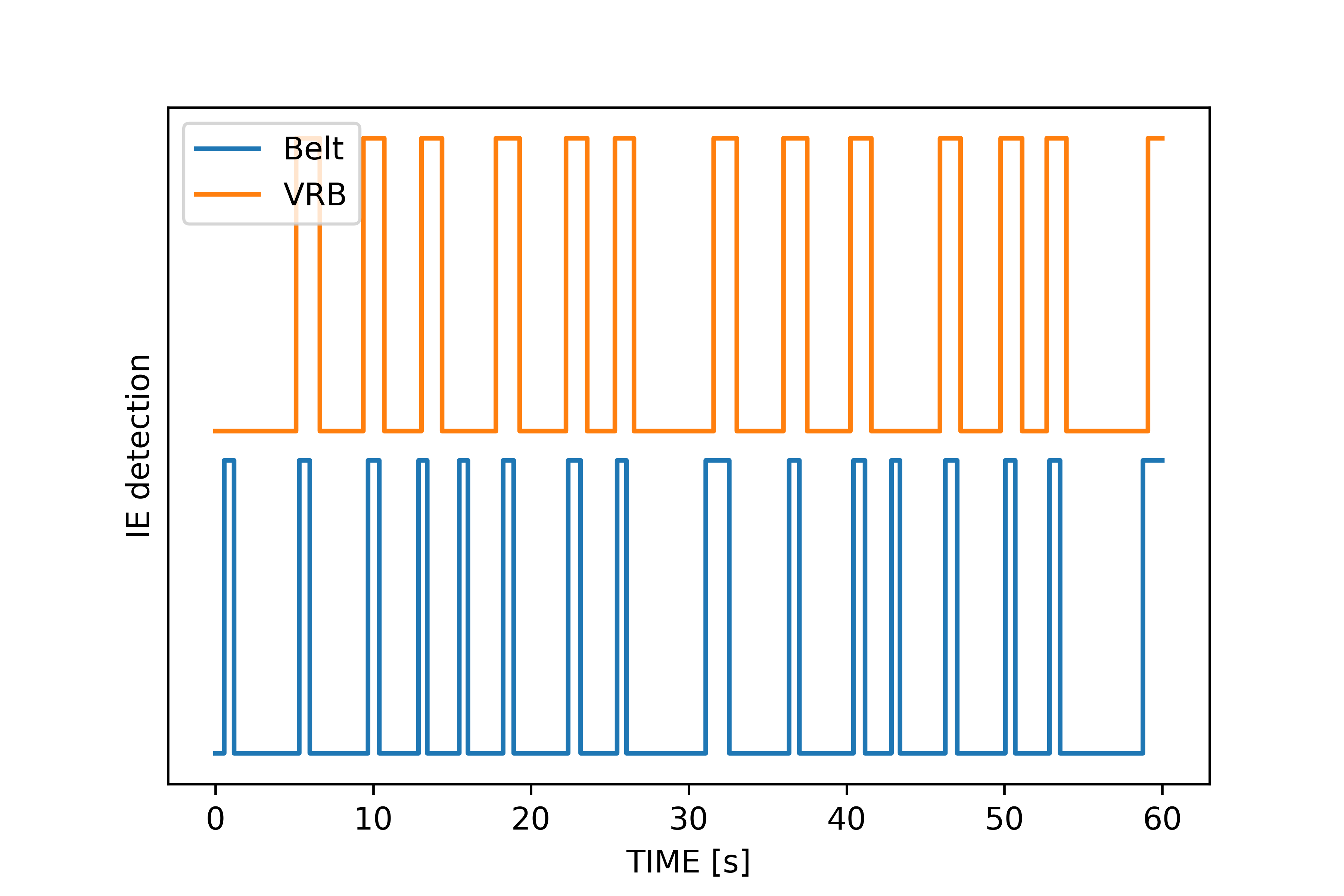}}\vspace{-8mm}
    \subfigure(c){\includegraphics[width=12cm]{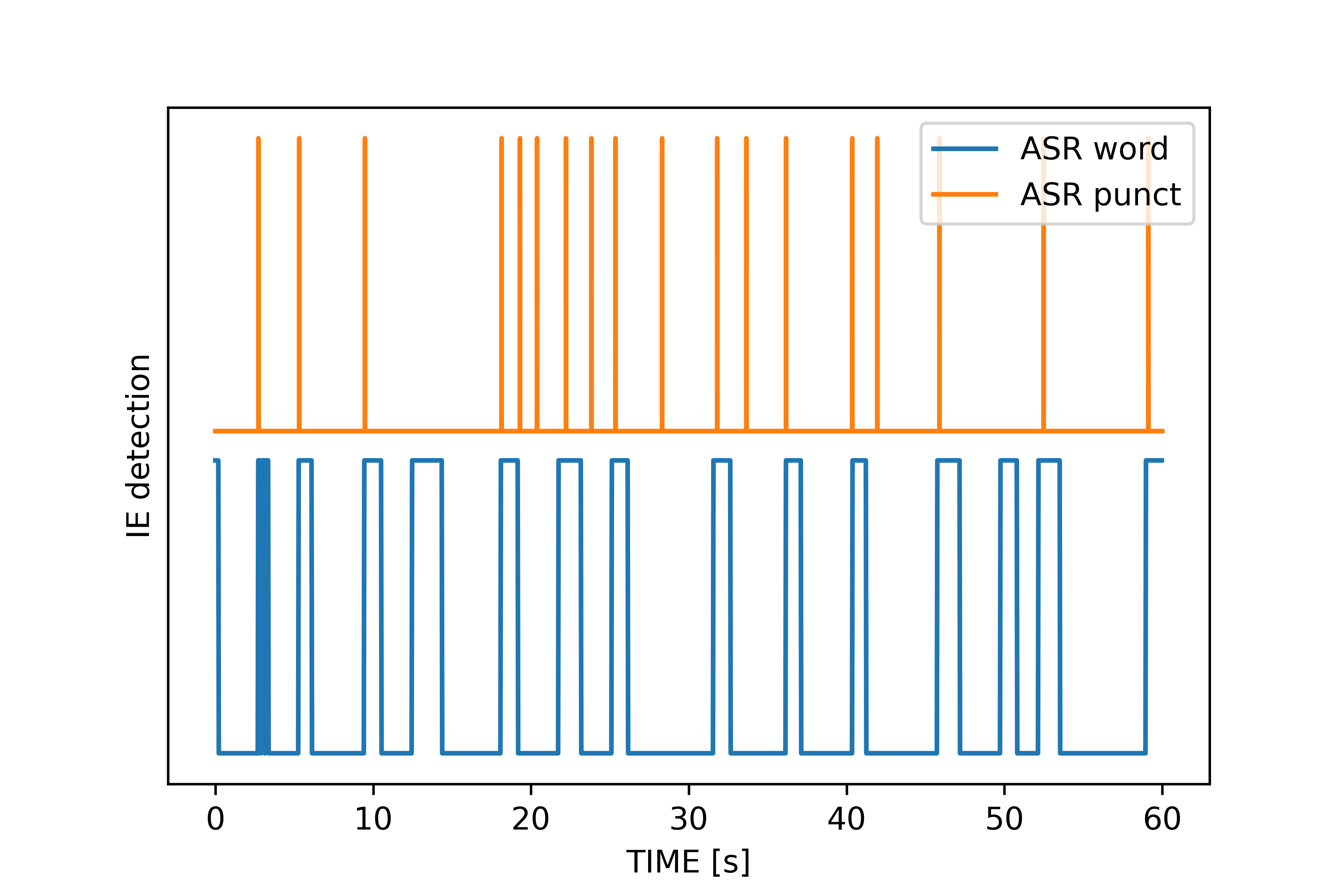}}\vspace{-2mm}  
    \caption{Examples of a) sensor belt signal and estimated VRB, b) detected IEs in the belt and VRB signals, and c) word pauses and punctuation points at the output of ASR.}
    \label{fig:waves}
\end{figure}

Fig\,\ref{fig:waves}a shows an example of a true sensor belt signal and an VRB estimate from the VRB model. The IE detections based are shown in Fig.\,\ref{fig:waves}b. Fig. Fig.\,\ref{fig:waves}c shows the long word pauses and grammatical stops based on the time-aligned ASR output. 

Based on the IE segments identified based on VRB, word pauses, or grammatical stops, we can compute performance metrics of a method in comparison to the {\it ground truth IEs} computed from the real sensor belt signal. True Positive detection is counted if an IE estimate temporally overlaps with the true IE and True Negative is counted when there are no IE estimates in regions between ground truth IEs. False Positives and Negatives are counts of IEs between ground truth IEs, and counts of missed ground truth IEs, respectively. 

The results of read "Rainbow" passage are shown in Table \ref{tab:read}. The overall performance, by the F1 score, is the highest, 0.85, for the VRB detections followed by ASR-word and ASR-punct. The performance of ASR-punct is significantly reduced by a high number of false negatives (fn). This is an interesting observation because it is in opposition with the observation of \cite{winkworth1994variability} that IEs occur, almost exclusively, in grammatically appropriate points. By the relatively high specificity, 0.72, it indeed seems to be the case that grammatical stops are often points of IEs. But, the high number of false negatives is a clear indication of ungrammatical breathing, that is, a large number of IEs in this read speech dataset occur in other parts of read content than close to grammatical stops. 

The ASR-word method based on long word pauses has a large number of both false positives and false negatives but overall performance with F1 score 0.74 is better than the ASR-punct method. Pauses in speech an IEs are indeed aligned, but not all long pauses are IEs.

\begin{table}[h]
\centering
\begin{tabular}{lrrrr}
\toprule
{} &  ASR-word &  ASR-punct &    VRB & VRBOLA \\
\midrule
tp          &    960  &     752  & 885  & 929\\
tn          &    985  &     762  & 866  & 915\\
fp          &    343  &     319  & 101  & 93\\
fn          &    347  &     568  & 290  & 232\\
Sensitivity &      0.73 &       0.57 &   0.75 & 0.80\\
Specificity &      0.74 &       0.70 &   0.90 & 0.91\\
F1-score    &      0.74 &       0.63 &   0.82 & 0.85\\
\bottomrule
\end{tabular}
\caption{Performance of IE detection in read speech}
\label{tab:read}
\end{table}

The performance of the three methods, see Table \ref{tab:spont}, have a similar trend also in the spontaneous speech dataset with the VRB method giving the highest overall performance. However, the performance of the ASR-word method is here significantly closer to VRB. One possible reason may be in the recording setup where the subject was instructed to talk much and the interviewer was just triggering more talk with brief questions. Many subjects were speaking in a very relaxed way and had relatively longer IEs than in read speech. In the ASR-punct one can see again increase in false negatives, that is, ungrammatical breathing but also an increased number of false positives. 

\begin{table}[h]
\centering
\begin{tabular}{lrrrr}
\toprule
{} &  ASR-word &  ASR-punct &    VRB & VRBOLA \\
\midrule
tp          &    513  &     370  & 512  & 532\\
tn          &    517  &     389  & 488  & 507\\
fp          &    269  &     312  & 136  & 135\\
fn          &    207  &     266  & 168  & 175\\
Sensitivity &      0.71 &       0.58 &   0.75 & 0.75\\
Specificity. &      0.66 &       0.55 &   0.78 & 0.79\\
F1-score          &      0.68 &       0.56 &   0.77 & 0.78\\
\bottomrule
\end{tabular}
\caption{Performance of IE detection in spontaneous speech}
\label{tab:spont}
\end{table}

The comparison of the conventional concatenated VRB processing, and the new VRBOLA method proposed in this paper shows a clear advantage for the proposed overlap-add processing. The VRBOLA method increases the computational requirements but it seems to increase the overall performance probably due to elimination of transients in frame borders. 

\section{Discussion}
Ingressive speech is rare in most languages and speech is largely phonetically controlled expiration. Consequently, inspiration takes place, almost exclusively, in pauses between words, or in nonverbal vocalization during gasping or yawning. The results of the experiment reported in this paper shows that detection of long pauses in speech can give a rough detection of inspiration events, IEs, in speech. However, not all pauses are used for inhaling and talkers are able to ventilate in very brief, less than 100ms, pauses between words. 

It is often suggested that IEs take mostly place in grammatically appropriate places in speech. In a study by Winksworth {\it et al.} \cite{winkworth1994variability} breathing in read speech is almost completely aligned with syntactic stops. The results of the current paper do not support this observation but nearly half of the IEs seem to take place in other parts of speech. In this paper we use the term ungrammatical breathing for this phenomenon and it was shown that it is common in both read and spontaneous speech. Obviously, not all paragraph, sentence, clause, and phrase boundaries are IEs. In addition, ambiguities in ASR and parsing, and ungrammatical speech may also lead to well-known errors, and spontaneous speech may have ungrammatical passages of words. 

Compared to the word pauses and grammatical stops, the best performance in the detection of IEs was with the Virtual Respiratory Belt, VRB, algorithm trained to model respiratory behavior of a talker from speech. The F1 score in the correct detection of IEs was here over 0.85 while in the methods based on word pauses or grammatical stops F1 score was below 0.75. 

The spontaneous speech in the current paper consists of speech fragments from a conversation but the actual turn-taking was not included in the analysis. Breathing in conversation is also influenced by the turn-taking behavior, see, e.g., \cite{wlodarczak_breathing_2020}, and in future work it would be interesting to explore the use of the talk of the other party so support IE detection. 

\section{Acknowledgements}

The authors wish to thank anonymous reviewers for their useful comments and [anon] for the financial support of this work. 

\balance

\bibliographystyle{IEEEtran}
\bibliography{pathological_speech,srikanth_refs}

\end{document}